\documentclass[aps,showpacs,preprint,preprintnumber,nofootinbib,amsmath,amssymb,
ascmac,bm,12pt]{revtex4}
\usepackage{bm}
\usepackage[dvipdfmx]{graphicx}
\usepackage[dvips]{color}

\begin{document}

\vspace*{0.5cm}
\title{
Kaluza-Klein vacuum multi-black holes in five-dimensions 
}
\author{
${}^{1,3}$Ken Matsuno\footnote{E-mail: matsuno@sci.osaka-cu.ac.jp},
${}^{1}$Hideki Ishihara\footnote{E-mail: ishihara@sci.osaka-cu.ac.jp},
${}^{2}$Masashi Kimura\footnote{E-mail: mkimura@yukawa.kyoto-u.ac.jp}
and
${}^{1}$Takamitsu Tatsuoka\footnote{E-mail: tatsuoka@sci.osaka-cu.ac.jp}
\bigskip
\bigskip
}
\affiliation{
${}^{1}$Department of Mathematics and Physics, Osaka City University, Sumiyoshi, 
Osaka 558-8585, Japan
\\
${}^{2}$Yukawa Institute for Theoretical Physics, Kyoto University, Kyoto 606-8502, 
Japan
\\
${}^{3}$Interdisciplinary Faculty of Science and Engineering, Shimane
University, Shimane 690-8504, Japan
\bigskip
\bigskip
}
\begin{abstract}
We investigate five-dimensional vacuum solutions 
which represent rotating multi-black holes in asymptotically Kaluza-Klein 
spacetimes. 
We show that multi-black holes rotate maximally along extra dimension, 
and stationary configurations in vacuum are achieved by the balance of the 
gravitational attraction force and repulsive force caused by the rotations 
of black holes. 
We also show that  each black hole can have the different topology of the 
lens space in addition to the spherical topology, 
and mass of black holes are quantized by the size of extra dimension and 
horizon topology. 
\end{abstract}

\preprint{OCU-PHYS-369}
\preprint{AP-GR-100}
\preprint{YITP-12-45}


\date{\today}
\maketitle

%%%%%%%%%%%%%%%%%%%%%%%%%%%%%%%%%%%%%%%%%%%%%%%%%%%%%%%%%%%%%%%%%%%%%%%%%%%%%%
\section{introduction}
%%%%%%%%%%%%%%%%%%%%%%%%%%%%%%%%%%%%%%%%%%%%%%%%%%%%%%%%%%%%%%%%%%%%%%%%%%%%%%

Kaluza-Klein black holes, which have compact extra dimensions, are  
interesting spacetimes in a class of higher-dimensional black holes. 
In particular, there exist a variety of exact solutions of 
five-dimensional squashed Kaluza-Klein black holes, which have 
squashed S$^3$ horizons 
\cite{Dobiasch:1981vh, Gibbons:1985ac, Gauntlett:2002nw, 
Gaiotto:2005gf, Ishihara:2005dp, 
Wang:2006nw, 
Yazadjiev:2006iv, Nakagawa:2008rm, Tomizawa:2008hw, Tomizawa:2008rh, Stelea:2008tt,  
Tomizawa:2008qr, Bena:2009ev, Tomizawa:2010xq, Mizoguchi:2011zj, 
Chen:2010ih, Nedkova:2011hx, Nedkova:2011aa, Tatsuoka:2011tx}.  
The solutions behave as fully five-dimensional black holes in the vicinity of 
the horizon, 
while they behave as four-dimensional black holes in the region far away 
from the horizon. 
The squashed Kaluza-Klein black holes asymptote to four-dimensional flat spacetimes 
with a twisted S$^1$ as a compactified extra dimension, 
and we can regard a series of these solutions as models of realistic 
higher-dimensional black holes.

In the Einstein-Maxwell theory in four dimensions, 
it is well known that extremal charged black holes 
can make multi-configurations \cite{Majumdar, Papapetrou}.
In five-dimensions, multi-black hole solutions are obtained both 
in asymptotically flat spacetimes 
\cite{Myers:1986rx, Breckenridge:1996is} 
and in asymptotically Kaluza-Klein spacetimes 
\cite{Myers:1986rx, Maeda:2006hd, Ishihara:2006iv, Matsuno:2008fn}. 
These solutions are achieved by the balance of gravitational attraction and 
electrical repulsion. 
Until now, any regular asymptotically flat vacuum multi-black hole solutions with spherical horizons 
in four or five dimensions have not been found \cite{Bunting:1987, Chrusciel:2005pa}.\footnote{
For non-spherical horizons in five dimensions, 
black Saturns and black di-rings are 
found as vacuum exact solutions \cite{Elvang:2007rd, Iguchi:2007is, Izumi:2007qx}.}

In contrast to the asymptotically flat case, in asymptotically Kaluza-Klein 
spacetimes, in this paper, 
we show that the metrics found by Clement \cite{IPUC-85-1} describe 
five-dimensional regular maximally rotating vacuum multi-black holes 
with a twisted S$^1$ as an extra dimension. 
Each black hole has a horizon with the topology of lens space 
$L(n; 1) =$ S$^3/\mathbb{Z}_n$. 
If the size of extra dimension $L$ is fixed, the regularity condition requires the 
quantization of black hole mass by $n L$.   

This paper is organized as follows. 
In Sec.\ref{solution}, we present explicit forms of solutions and constraints between parameters of the solutions.   
Section \ref{Properties}  
is devoted to an investigation of conserved charges, asymptotic structures 
of the solutions, and the regularity at the horizons. 
We conclude our studies with discussions in Sec.\ref{Discussions}.

%%%%%%%%%%%%%%%%%%%%%%%%%%%%%%%%%%%%%%%%%%%%%%%%%%%%%%%%%%%%%%%%%%%%%%%%%%%%%%
\section{solutions}\label{solution}
%%%%%%%%%%%%%%%%%%%%%%%%%%%%%%%%%%%%%%%%%%%%%%%%%%%%%%%%%%%%%%%%%%%%%%%%%%%%%%
We consider rotating multi-black hole solutions satisfying 
the five-dimensional vacuum Einstein equation, $R_{\mu \nu } = 0$. 
The metric is written as 
\begin{align}
\label{mET1}
	ds^2 = - H^{-2} dt^2 + H^2 (dx^2+dy^2+dz^2) 
		+ 2 \left[ \left(H^{-1} -1 \right) dt + \frac{L}{2 \sqrt 2} d\psi 
		+ \bm \omega \right]^2 ,
\end{align}
where 
\begin{align}
	H &= 1 + \sum_i \frac{m_i}{|\bm R - \bm R _i|} 
\label{FUNCH} 
\end{align}
is the harmonic function on the three-dimensional Euclid space 
with point sources located at $\bm R = \bm R_i := (x_i, y_i, z_i)$.   
The 1-form $\bm \omega $, which is determined by 
\begin{align}
\nabla \times \bm \omega = \nabla H ,
\end{align}
has the explicit form 
\begin{align}
\bm \omega =  
\sum_i m_i \frac{z-z_i}{|\bm R - \bm R _i|}
\frac{(x-x_i)dy-(y-y_i)dx}{(x-x_i)^2+(y-y_i)^2} . 
\label{one-form}
\end{align}
In the expressions \eqref{mET1}-\eqref{one-form}, $m_i$ and $L$ are constants. 
The solutions \eqref{mET1} with \eqref{FUNCH} and \eqref{one-form} can be obtained 
by uplifting the four-dimensional equally charged dyonic Majumdar-Papapetrou 
solutions with a constant dilaton field 
to the five-dimensional spacetimes \cite{IPUC-85-1}. 
(See Appendix \ref{KKRED} for detail discussion.)\footnote{
In the single black hole case, $m_1 = m$ and $m_i = 0 ~(i \geq 2)$, 
the solution \eqref{mET1} coincides with 
an extremally rotating vacuum squashed Kaluza-Klein black hole solution 
with a degenerate horizon \cite{Gibbons:1985ac}. 
The solution \eqref{mET1} was also obtained 
in the context of the ten-dimensional $N = 1$ supergravity \cite{Khuri:1995xq}. 
However, to the best our knowledge, 
properties of the solution \eqref{mET1} 
like asymptotic structures and a smoothness of horizons have not been discussed.}  
As will be shown later, $\bm R = \bm R_i$ are black hole horizons.

From the requirements for the absence of naked singularity 
on and outside the black hole horizons, 
the parameters are restricted to the range
\begin{align}\label{PARAREGS}
	m _i > 0 . 
\end{align}
We will see later that the regularity of horizons requires 
the parameters $m_i$ to be quantized by 
the size of the compactified dimension $L$ at infinity in the form
\begin{align}\label{QUANTIZE}
	m _i = \frac{n _i L}{2 \sqrt 2} , 
\end{align}
where $n_i$ are the natural numbers.

%%%%%%%%%%%%%%%%%%%%%%%%%%%%%%%%%%%%%%%%%%%%%%%%%%%%%%%%%%%%%%%%%%%%%%%%%%%%%%

%%%%%%%%%%%%%%%%%%%%%%%%%%%%%%%%%%%%%%%%%%%%%%%%%%%%%%%%%%%%%%%%%%%%%%%%%%%%%%
\section{Properties}
\label{Properties}
%%%%%%%%%%%%%%%%%%%%%%%%%%%%%%%%%%%%%%%%%%%%%%%%%%%%%%%%%%%%%%%%%%%%%%%%%%%%%%
%%%%%%%%%%%%%%%%%%%%%%%%%%%%%%%%%%%%%%%%%%%%%%%%%%%%%%%%%%%%%%%%%%%%%%%%%%%%%%
\subsection{Basic properties }
%%%%%%%%%%%%%%%%%%%%%%%%%%%%%%%%%%%%%%%%%%%%%%%%%%%%%%%%%%%%%%%%%%%%%%%%%%%%%%
It is clear that the metric \eqref{mET1} admits 
two Killing vector fields, 
\begin{align}
\xi_{(t)} = \partial / \partial t \quad 
\mbox{and} \quad
\xi_{(\psi)} = \partial / \partial \psi. 
\end{align}
Timelike Killing vectors which are timelike at 
infinity are not unique because the Killing vector 
along the compact extra dimension 
$\xi_{(\psi)}$ has a finite norm at the infinity. 
This fact is quite different from the asymptotically flat case, where only $\xi_{(t)}$ 
is timelike at the infinity. 
Fortunately, however, we can select out the timelike Killing vector 
$\xi_{(t)}$ that is hypersurface orthogonal at the infinity among them. 
We define the Komar mass  $M$ associated with 
$\xi_{(t)}$, and obtain as 
\begin{align}
	M &= \frac{-3}{32\pi}\int_\infty dS_{\mu\nu}\nabla^\mu \xi_{(t)}^\nu
		= \frac{3 L \sum_i m_i}{4 \pi} \mathcal A_{\rm S ^3} ,
\label{mASS}
\end{align}
where $\mathcal A_{\rm S ^3}$ denotes the area of a unit S$^3$. 
We also obtain the angular momentum $J_{\psi}$ 
associated with the spacelike Killing vector 
$\xi_{(\psi)}$ 
as
\begin{align}
 J^{\psi} &= \frac{1}{16\pi}\int_\infty dS_{\mu\nu}
		\nabla^\mu \xi_{(\psi)}^\nu
	= \frac{L ^2 \sum_i m_i}{4 \sqrt 2 \pi} \mathcal A_{\rm S ^3} .
\label{ANGmOm}
\end{align}
We see that the spacetime \eqref{mET1} is rotating along 
the extra dimension. 
Because the present solutions are vacuum solutions, the mass and angular 
momentum can be assigned to each black hole 
\begin{align}
	M_i 
		= \frac{3 L m_i}{4 \pi} \mathcal A_{\rm S ^3} ,
\label{mASS_each}
\\
	J^{\psi}_i 
	= \frac{L ^2 m_i}{4 \sqrt 2 \pi} \mathcal A_{\rm S ^3} ,
\label{ANGmOm_each}
\end{align}
by taking integral for the closed 
surface surrounding each black hole. 

Substituting \eqref{QUANTIZE} into \eqref{mASS_each} and \eqref{ANGmOm_each}, 
we obtain a relation between the mass and the angular momentum as  
\begin{align}
	(J_i^{\psi})^{2} = \frac{2 \sqrt 2}{27 \pi n_i} M_i^3 . 
\end{align}
This relation means that each black hole is maximally rotating. 
We can also obtain the relation 
\begin{align}
\frac{J_i^\psi}{M_i} = \frac{L}{3 \sqrt 2} . 
\end{align}
Total mass and total angular momentum are given by the summations, 
\begin{align}
	M=\sum_i M_i, 
\quad
	J^{\psi}=\sum_i J^{\psi}_i, 
\end{align}
which satisfy the condition
\begin{align}
	(J^{\psi})^{2} = \frac{2 \sqrt 2}{27 \pi n} M^3, 
\end{align}
where $n = \sum_i n_i$.

With respect to the timelike Killing vector $\xi_{(t)}$, we define 
the ergosurfaces where the Killing vector becomes null, i.e., 
\begin{align}\label{ERGOEQ}
	g_{tt} = \left( H^{-1} -2 \right)^2 -2 = 0 . 
\end{align}
%%%%%%%%%%%%%%%%%%%%%%%%%%%%%%%%%%%%%%%%%%%%%%%%%%%%%%%%%%%%%%%%%%%%%%%%%%%%%%
In the single black hole case, $m_1 = m$ and $m_i = 0 ~(i \geq 2)$, 
the equation \eqref{ERGOEQ} reduces to 
\begin{align}
	g_{tt}= \frac{2 m^2 - R^2}{(R + m)^2} = 0 .
\end{align}
Then, the ergosurface exists at $R=\sqrt{2}m$. 
In general, since $g_{tt}(\bm R = \bm R_i) = 2 > 0$ and $g_{tt}(\infty) = -1 < 0$ 
for the range of parameters \eqref{PARAREGS}, 
there always exist ergoregions 
around the black hole horizons.
%%%%%%%%%%%%%%%%%%%%%%%%%%%%%%%%%%%%%%%%%%%%%%%%%%%%%%%%%%%%%%%%%%%%%%%%%%%%%%
It depends on the configuration of point sources whether the ergoregions are 
connected or not \cite{Matsuno:2008fn}.

%%%%%%%%%%%%%%%%%%%%%%%%%%%%%%%%%%%%%%%%%%%%%%%%%%%%%%%%%%%%%%%%%%%%%%%%%%%%%%
\subsection{Asymptotic structure }
%%%%%%%%%%%%%%%%%%%%%%%%%%%%%%%%%%%%%%%%%%%%%%%%%%%%%%%%%%%%%%%%%%%%%%%%%%%%%%
We assume that point sources exist in a bounded domain. 
In the far region from the domain, 
the harmonic function $H$ and the 1-form $\bm \omega$ behave as  
\begin{align}
	H &\simeq 1 + \frac{\sum _i m_i}{R} + O\left( R ^{-2} \right) ,
\\ 
	\bm \omega &\simeq \left( \sum _i m_i \right) 
		\cos \theta d \phi + O\left( R ^{-1} \right) .
\end{align}
Then, using \eqref{QUANTIZE}, we see that the metric behaves as
\begin{align}
	ds^2 \simeq - \left(1+\frac{m}{R}\right)^{-2} dt ^2 
		+ \left(1+\frac{m}{R}\right)^2 \left(dR^2 + R^2 d\Omega_{\rm S ^2}^2\right) 
	+ \frac{n ^2 L ^2}{4} 
		\left(-\frac{dt}{R} +\frac{d\psi}{n}+\cos \theta d \phi \right) ^2 , 
\end{align}
where $d\Omega _{\rm S ^2} ^2 = d\theta ^2 + \sin^2\theta d\phi^2$ 
denotes the metric of the unit two-sphere, 
and $m = \sum_i m_i$. 
The metric behaves as a single extremely rotating Kaluza-Klein 
black hole \cite{Dobiasch:1981vh, Gibbons:1985ac}.
In the limit $R\to \infty$, 
we see that the metric approaches as 
\begin{align}
	ds^2 \to - dt ^2 +  dR^2 + R^2 d\Omega _{\rm S ^2} ^2 
	+ \frac{n ^2 L ^2}{4} \left( \frac{d\psi}{n} + \cos \theta d \phi \right) ^2 . 
\label{asympt_metric}
\end{align}
The asymptotic structure of the spacetime \eqref{mET1} is asymptotically 
locally flat, i.e., 
the metric asymptotes to a twisted constant S$^1$ fiber bundle over 
the four-dimensional Minkowski spacetime, 
and the spatial infinity has the structure of an S$^1$ bundle over an S$^2$ 
such that it is the lens space 
$L(n; 1) =$ S$^3/\mathbb{Z}_n$ \cite{Ishihara:2006iv,Matsuno:2008fn}. 
We see that the size of a twisted S$^1$ fiber as an extra dimension takes 
the constant value $L$ everywhere.

%%%%%%%%%%%%%%%%%%%%%%%%%%%%%%%%%%%%%%%%%%%%%%%%%%%%%%%%%%%%%%%%%%%%%%%%%%%%%%
\subsection{Near horizon}
%%%%%%%%%%%%%%%%%%%%%%%%%%%%%%%%%%%%%%%%%%%%%%%%%%%%%%%%%%%%%%%%%%%%%%%%%%%%%%

For simplicity, we restrict ourselves to the cases of two-black holes, 
i.e., $m _i = 0 ~(i \geq 3)$.
Without loss of generality, we can put the locations of two point sources as 
$\bm R_1 = (0, 0, 0)$ and $\bm R_2 = (0, 0, a)$, 
where the constant $a$ denotes the separation between two black holes.

In this case, the metric is  
\begin{align}\label{mET2}
	ds^2 = - H^{-2} dt^2 + H^2 
	\left( dR^2 + R^2 d\Omega _{\rm S ^2} ^2 \right) 
		+ 2 \left[ \left(H^{-1} -1 \right) dt + \frac{L}{2 \sqrt 2} d\psi 
		+ \bm \omega \right]^2 ,
\end{align}
where $H$ and $\bm \omega$ are given by 
\begin{align}
	H &= 1+ \frac{m_1}{R} 
		+ \frac{m_2}{\sqrt{R^2 + a^2 - 2 a R \cos \theta}} ,
\label{harmonics_2}
\\
	\bm \omega &= \left( 
		m_1 \cos \theta + m_2 \frac{R \cos \theta - a }
		{\sqrt{R^2 + a^2 - 2 a R \cos \theta}} \right) d\phi,
\label{form_2}
\end{align}
respectively. 
The coordinates run the ranges of 
$-\infty < t < \infty , ~0 < R < \infty,~0 \leq \theta \leq\pi,~ \quad
0 \leq \phi \leq 2\pi $, and $0 \leq \psi \leq 4\pi $.

In the coordinate system $(t, R, \theta , \phi , \psi )$,
the metric \eqref{mET2} diverges at the locations of two point sources, i.e., 
$\bm R = \bm R_1 ~(R=0)$ and $\bm R = \bm R_2 ~(R=a, \theta=0)$.

In order to remove apparent divergences 
at $R = 0$,
we introduce new coordinates $(v, \psi' )$ such that 
\begin{align}
	dv &= dt + H^2 dR + W d\theta ,
\\
	d\psi'
	&= d\psi - \frac{2 \sqrt 2}{L}\left(dt + H dR + V d\theta 
		\right) , 
\end{align}
where the functions $W$ and $V$ are given by 
\begin{align}
	W \left(R, \theta\right)
		&= \int dR \frac{\partial}{\partial \theta}\left( H^2 \right) ,
\label{FUNCW}
\\
	V \left(R, \theta\right)&= \int dR \frac{\partial}{\partial \theta} H , 
\label{FUNCV}
\end{align}
respectively. 
Then, the metric \eqref{mET2} takes the form of 
\begin{align}
	ds^2 = & - H^{-2} \left(dv - W d\theta \right)^2 
		+ 2 dR \left(dv - W d\theta \right)
		+ H^2 R^2 d\Omega _{\rm S ^2} ^2
\notag \\
		&\hspace{7mm} + 2 \left[ 
			H^{-1} dv + \bm \omega 
			+ \left(V - H^{-1}W \right) d\theta + \frac{L}{2 \sqrt 2} d\psi'
		\right]^2 . 
\label{mET3}
\end{align}

In the neighborhood of $R = 0$, the functions $H,~W, ~V$, and the 1-form 
$\bm \omega$ 
can be expanded by power series in $R$, and leading orders are 
\begin{align}
	H &\simeq \frac{m_1}{R} 
		+ {\cal O}(1) ,
\label{EXPANDH}
\\
	W &\simeq - \frac{2 m_1 m_2 \sin \theta}{a^2} R 
		+{\mathcal O}(R^2), 
\label{EXPANDW}
\\
	V &\simeq 
		- \frac{m_2 \sin \theta}{2a^2} R^2 	+{\cal O}(R^3), 
\label{EXPANDV}
\\
	\bm \omega &\simeq \left( m_1 \cos \theta - m_2 
	+{\cal O}(R^2)
	\right)d\phi ,
\label{EXPANDOmEGA}
\end{align}
where we have chosen the integral constants of $W$ and $V$ 
given by \eqref{FUNCW} and \eqref{FUNCV} suitably. 
Then, near $R = 0$, the metric \eqref{mET3} behaves as 
\begin{align}
	ds^2 \simeq & \frac{R^2}{m_1 ^2} dv^2 + 2 dv dR 
		+ m_1 ^2 \left[ d\Omega _{\rm S ^2} ^2 
	+ 2 \left(\frac{L}{2\sqrt{2}m_1}d\psi'' 
		+ \cos \theta d\phi \right)^2 \right] 
\cr
		& + 4 R \left[ 
			\frac{m_1 m_2 \sin \theta }{a^2} dR d\theta 
			+ \left( dv + \frac{2 m_1 m_2 \sin \theta }{a^2} R d\theta \right) 
			\left( \frac{L}{2\sqrt{2}m_1}d\psi'' + \cos \theta d\phi \right) 
	\right] \cr 
	& + {\mathcal O}(R^3) ,
\label{near_horison_metric}
\end{align}
where we have used 
\begin{align}
	d\psi'' = d\psi' - \frac{2\sqrt 2}{L} m _2 d\phi . 
\end{align}
If the factor $2\sqrt{2}m_1/L$ is a natural number, say $n_1$, 
the induced metric on the three-dimensional spatial cross section of 
$R = 0$ with a time slice is  
\begin{align}\label{INDmET}
	ds^2 |_{R=0} 
	= \frac{n_1^2 L^2}{8} \left[ d\Omega _{\rm S ^2} ^2
		+ 2 \left( \frac{d\psi''}{n_1} + \cos \theta d\phi \right)^2 \right] . 
\end{align}
That is, if the mass quantization condition \eqref{QUANTIZE} holds, the $R=0$ 
surface admits the smooth metric of the squashed lens space 
$L(n_1;1)=$ S$^3/\mathbb{Z}_{n_1}$. 
The area of the surface is  
\begin{align}
	\mathcal A|_{R=0} 
		= \frac{n_1^2 L^3}{2} \mathcal A_{\rm S ^3} . 
\end{align}

Under the condition \eqref{QUANTIZE}, 
we see that $R=0$ is a null surface where 
the metric \eqref{mET3} is regular and each component is 
an analytic function of $R$.
Therefore the metric \eqref{mET3} gives analytic extension 
across the surface $R = 0$. 
By the same discussion, we see that 
the metric \eqref{mET2} also admits analytic extension across 
the surface $\bm R = \bm R _2$. 

We also see that $\eta=\partial_v$ is a Killing vector field 
that becomes null at $R=0$. 
Furthermore, $\eta$ is hypersurface orthogonal to the surface $R=0$, i.e., 
$\eta_{ \mu} dx^\mu = g_{vR} dR = dR $ there. 
These mean that the null hypersurface $R=0$ is a Killing horizon. 
Similarly,  $\bm R = \bm R _2$ is also a Killing horizon. 
Hence, we can see that 
the solutions \eqref{mET2} with \eqref{harmonics_2} and \eqref{form_2} describe 
Kaluza-Klein multi-black holes, 
which have smooth Killing horizons without singularity on and outside 
the black hole horizons. 
The topology of each black hole horizon is  
the lens spaces $L(n_i;1)$.   
Since the  $\phi$-$\psi$ part of the metric is positive definite, 
it is clear that no closed timelike curve exists. 
Near each horizon limit, the metric \eqref{mET2} 
approaches the $L(n_i;1)$ bundle 
over the AdS$_2$ space at the horizon \cite{Reall:2002bh,Kunduri:2007vf}.

%%%%%%%%%%%%%%%%%%%%%%%%%%%%%%%%%%%%%%%%%%%%%%%%%%%%%%%%%%%%%%%%%%%%%%%%%%%%%%

%%%%%%%%%%%%%%%%%%%%%%%%%%%%%%%%%%%%%%%%%%%%%%%%%%%%%%%%%%%%%%%%%%%%%%%%%%%%%%
\section{summary and discussions}
\label{Discussions}
%%%%%%%%%%%%%%%%%%%%%%%%%%%%%%%%%%%%%%%%%%%%%%%%%%%%%%%%%%%%%%%%%%%%%%%%%%%%%%
We have investigated extremely rotating Kaluza-Klein multi-black hole solutions 
in the five-dimensional pure Einstein theory given by the metric \eqref{mET1} 
with \eqref{FUNCH} and \eqref{one-form}.  
The metric asymptotes to the effectively four-dimensional 
spacetime and the size of the compactified extra dimension takes the constant 
value everywhere.  
We have shown that 
each black hole has a smooth horizon and its topology is the lens space.  
Furthermore, 
the mass and the angular momentum of the black hole satisfy the extremality 
condition and  
the horizon size of each black hole is quantized by the size of the compactified 
dimension. 
To sum up, 
the exact solutions describe five-dimensional regular vacuum rotating Kaluza-Klein 
multi-black holes.

In the solutions, for each black hole, 
the ratio of the mass and the angular momentum is determined rigidly by 
a value of order of unity. 
We can interpret that this comes from the force balance between 
gravitational force and spin-spin repulsive force between black holes. 
This corresponds to the balance of 
gravitational force and Coulomb repulsive force after Kaluza-Klein reduction.
Furthermore, the black hole mass should be quantized by the size of 
extra dimension $L$ from the regularity of the horizon. 
The minimum size of black hole, which is comparable to $L$, exists. 
Then, we cannot get asymptotically flat solutions from the present solutions 
by taking a limit $L \to \infty$ keeping the black hole mass constant. 
This is consistent with the fact that no vacuum multi-black hole has been found 
in five-dimensional asymptotically flat spacetime. 
The asymptotic structure of the solution with a compact dimension affects 
the existence of the multi black holes. 
Whether more general solutions exist or not is open question.

%%%%%%%%%%%%%%%%%%%%%%%%%%%%%%%%%%%%%%%%%%%%%%%%%%%%%%%%%%%%%%%%%%%%%%%%%%%%%%
\section*{Acknowledgments}
%%%%%%%%%%%%%%%%%%%%%%%%%%%%%%%%%%%%%%%%%%%%%%%%%%%%%%%%%%%%%%%%%%%%%%%%%%%%%%
This work is supported by the Grant-in-Aid for Scientific Research No.19540305. 
MK is supported by the JSPS Grant-in-Aid for Scientific Research No.11J02182.

%%%%%%%%%%%%%%%%%%%%%%%%%%%%%%%%%%%%%%%%%%
\appendix
%%%%%%%%%%%%%%%%%%%%%%%%%%%%%%%%%%%%%%%%%%%%%%%%%%%%%%%%%%%%%%%%%%%%%%%%%%%%%%
\section{Kaluza-Klein reduction}\label{KKRED}
%%%%%%%%%%%%%%%%%%%%%%%%%%%%%%%%%%%%%%%%%%%%%%%%%%%%%%%%%%%%%%%%%%%%%%%%%%%%%%
We consider the Kaluza-Klein parametrization of a five-dimensional metric, 
\begin{align}
ds^2 = e^{- \Phi /3} ds_{(4)} ^{2} 
+ e^{2\Phi /3} \left( \frac{L}{2} d\psi + \bm A \right)^2 
\end{align}
which admits the Killing vector field $\partial_\psi$, 
where $ds_{(4)} ^{2} , ~\bm F = d \bm A$, and $\Phi$ are identified with    
the four-dimensional metric, the four-dimensional Maxwell field, 
and the dilaton field, respectively.   
The metric $ds_{(4)} ^{2}$ and two fields $\bm F, ~\Phi$ satisfy the field equations of the four-dimensional Einstein-Maxwell-dilaton theory with the action
\begin{align}
S = \frac{1}{16\pi G_4} \int d^4 x \sqrt{-g_{(4)}} 
\left(
	R_{(4)} -\frac{1}{6} \partial_\mu \Phi \partial^\mu \Phi 
	-\frac{e^\Phi}{4} F_{\mu\nu} F^{\mu\nu} 
\right) .
\end{align}

After the Kaluza-Klein reduction of the metric \eqref{mET1} 
with respect to the Killing vector field $\partial_\psi$,  
we find  
\begin{align}
 ds_{(4)}^2 &= - H^{-2} dt^2 + H^2 (dx^2+dy^2+dz^2) ,
\label{4DmET}
\\
 \bm F &= \sqrt 2 d \left(H^{-1} dt + \bm \omega \right) ,
\label{4DmAXWELL}
\\
 \Phi &= \text{const.} , 
\label{4DDILATON} 
\end{align}
where the function $H$ is given by \eqref{FUNCH} and 
the four-dimensional Maxwell field $\bm F$ satisfies 
the relation $F_{\mu\nu} F^{\mu\nu} = 0$.  
We see that \eqref{4DmET}-\eqref{4DDILATON} coincide with 
the four-dimensional equal charged dyonic Majumdar-Papapetrou solution 
with a constant dilaton field.  

%%%%%%%%%%%%%%%%%%%%%%%%%%%%%%%%%%%%%%%%%%%%%%%%%%%%%%%%%%%%%%%%%


\begin{thebibliography}{999}
%%%%%%%%%%%%%%%%%%%%%%%%%%%%%%%%%%%%%%%%%%%%%%%%%%%%%%%%%%%%%%%%%
%%%%%%%%%%%%%%%%%%%%%%%%%%%%%%%%%%%%%%%%%%%%%%%%%%%%%%%%%%%%%%%%%


%%%%%%%%%%%%%%%%%%%% SqKK %%%%%%%%%%%%%%%%%%%%%%%%%%%%%%%%
%\cite{Dobiasch:1981vh}
\bibitem{Dobiasch:1981vh}
  P.~Dobiasch and D.~Maison,
  %``Stationary, Spherically Symmetric Solutions Of Jordan's Unified Theory Of
  %Gravity And Electromagnetism,''
  Gen.\ Rel.\ Grav.\  {\bf 14}, 231 (1982).
  %%CITATION = GRGVA,14,231;%%

%\cite{Gibbons:1985ac}
\bibitem{Gibbons:1985ac}
  G.~W.~Gibbons and D.~L.~Wiltshire,
  %``Black Holes In Kaluza-Klein Theory,''
  Annals Phys.\  {\bf 167}, 201 (1986), 
  [Erratum-ibid.\  {\bf 176}, 393 (1987)].
  %%CITATION = APNYA,167,201;%%
  
%\cite{Gauntlett:2002nw}
\bibitem{Gauntlett:2002nw}
  J.~P.~Gauntlett, J.~B.~Gutowski, C.~M.~Hull, S.~Pakis and H.~S.~Reall,
  %``All supersymmetric solutions of minimal supergravity in five dimensions,''
  Class.\ Quant.\ Grav.\  {\bf 20}, 4587 (2003). 
  [arXiv:hep-th/0209114].
  %%CITATION = CQGRD,20,4587;%%
  
%\cite{Gaiotto:2005gf}
\bibitem{Gaiotto:2005gf}
  D.~Gaiotto, A.~Strominger, X.~Yin,
  %``New connections between 4-D and 5-D black holes,''
  JHEP {\bf 0602}, 024 (2006). 
  [hep-th/0503217].    

%\cite{Ishihara:2005dp}
\bibitem{Ishihara:2005dp}
  H.~Ishihara and K.~Matsuno,
  %``Kaluza-Klein black holes with squashed horizons,''
  Prog.\ Theor.\ Phys.\  {\bf 116}, 417 (2006). 
  [arXiv:hep-th/0510094].
  %%CITATION = PTPKA,116,417;%%

%\cite{Wang:2006nw}
\bibitem{Wang:2006nw} 
  T.~Wang,
  %``A Rotating Kaluza-Klein black hole with squashed horizons,''
  Nucl.\ Phys.\ B {\bf 756}, 86 (2006). 
  [hep-th/0605048].

%\cite{Yazadjiev:2006iv}
\bibitem{Yazadjiev:2006iv}
  S.~S.~Yazadjiev,
  %``Dilaton black holes with squashed horizons and their thermodynamics,''
  Phys.\ Rev.\  {\bf D74}, 024022 (2006). 
  [hep-th/0605271].  

%\cite{Nakagawa:2008rm}
\bibitem{Nakagawa:2008rm}
  T.~Nakagawa, H.~Ishihara, K.~Matsuno and S.~Tomizawa,
%  ``Charged Rotating Kaluza-Klein Black Holes in Five Dimensions,''
  Phys.\ Rev.\  D {\bf 77}, 044040 (2008). 
  [arXiv:0801.0164 [hep-th]].
  %%CITATION = PHRVA,D77,044040;%%

 %\cite{Tomizawa:2008hw}
\bibitem{Tomizawa:2008hw}
  S.~Tomizawa, H.~Ishihara, K.~Matsuno and T.~Nakagawa,
  %``Squashed Kerr-Godel Black Holes - Kaluza-Klein Black Holes with Rotations
  %of Black Hole and Universe -,''
  Prog.\ Theor.\ Phys.\  {\bf 121}, 823 (2009). 
  [arXiv:0803.3873 [hep-th]].
  %%CITATION = PTPKA,121,823;%%
  
%\cite{Tomizawa:2008rh}
\bibitem{Tomizawa:2008rh}
  S.~Tomizawa and A.~Ishibashi,
  %``Charged Black Holes in a Rotating Gross-Perry-Sorkin Monopole Background,''
  Class.\ Quant.\ Grav.\  {\bf 25}, 245007 (2008). 
  [arXiv:0807.1564 [hep-th]].
  %%CITATION = CQGRD,25,245007;%%

  %\cite{Stelea:2008tt}
\bibitem{Stelea:2008tt}
  C.~Stelea, K.~Schleich and D.~Witt,
 %``On squashed black holes in Godel universes,''
  Phys.\ Rev.\  D {\bf 78}, 124006 (2008). 
  [arXiv:0807.4338 [hep-th]].
  %%CITATION = PHRVA,D78,124006;%%  
  
%\cite{Tomizawa:2008qr}
\bibitem{Tomizawa:2008qr}
  S.~Tomizawa, Y.~Yasui, Y.~Morisawa,
  %``Charged Rotating Kaluza-Klein Black Holes Generated by G2(2) Transformation,''
  Class.\ Quant.\ Grav.\  {\bf 26}, 145006 (2009). 
  [arXiv:0809.2001 [hep-th]].  
  
%\cite{Bena:2009ev}
\bibitem{Bena:2009ev}
  I.~Bena, G.~Dall'Agata, S.~Giusto, C.~Ruef, N.~P.~Warner,
  %``Non-BPS Black Rings and Black Holes in Taub-NUT,''
  JHEP {\bf 0906}, 015 (2009). 
  [arXiv:0902.4526 [hep-th]]. 

%\cite{Tomizawa:2010xq}
\bibitem{Tomizawa:2010xq}
  S.~Tomizawa,
  %``Compactified black holes in five-dimensional $U(1)^3$ ungauged supergravity,''  
  arXiv:1009.3568 [hep-th]. 

%\cite{Mizoguchi:2011zj}
\bibitem{Mizoguchi:2011zj}
  S.~'y.~Mizoguchi, S.~Tomizawa,
  %``New approach to solution generation using SL(2,R)-duality of a dimensionally reduced space in five-dimensional minimal supergravity and new black holes,''
  Phys.\ Rev.\  {\bf D84}, 104009 (2011). 
  [arXiv:1106.3165 [hep-th]].
  
%\cite{Chen:2010ih}
\bibitem{Chen:2010ih} 
  Y.~Chen and E.~Teo,
  %``Black holes on gravitational instantons,''
  Nucl.\ Phys.\ B {\bf 850}, 253 (2011). 
  [arXiv:1011.6464 [hep-th]].
  
%\cite{Nedkova:2011hx}
\bibitem{Nedkova:2011hx} 
  P.~G.~Nedkova and S.~S.~Yazadjiev,
  %``On the Thermodynamics of 5D Black Holes on ALF Gravitational Instantons,''
  Phys.\ Rev.\ D {\bf 84}, 124040 (2011). 
  [arXiv:1109.2838 [hep-th]].

%\cite{Nedkova:2011aa}
\bibitem{Nedkova:2011aa} 
  P.~G.~Nedkova and S.~S.~Yazadjiev,
  %``Magnetized Black Hole on Taub-Nut Instanton,''
  Phys.\ Rev.\ D {\bf 85}, 064021 (2012). 
  [arXiv:1112.3326 [hep-th]].

  
%%%%%%%%%%%%%%%%%%%% SqKK %%%%%%%%%%%%%%%%%%%%%%%%%%%%%%%%

%%%%%%%%%%%%%%%%%%%%%%%%%%%%%%%%%%%%%%%%%%%%%%%%%%%%%%%%%%%
%\cite{Tatsuoka:2011tx}
\bibitem{Tatsuoka:2011tx} 
  T.~Tatsuoka, H.~Ishihara, M.~Kimura and K.~Matsuno,
  %``Extremal Charged Black Holes with a Twisted Extra Dimension,''
  Phys.\ Rev.\ D {\bf 85}, 044006 (2012). 
  [arXiv:1110.6731 [hep-th]].
%%%%%%%%%%%%%%%%%%%%%%%%%%%%%%%%%%%%%%%%%%%%%%%%%%%%%%%%%%%

\bibitem{Majumdar}
S.D.Majumdar, Phys. Rev.  {\bf 72}, 390 (1947). 

\bibitem{Papapetrou}
A. Papapetrou, Proc. R. Ir. Acad. Sect. A 51, 191 (1947). 

%%%%%%%%%%%%%%%%%%%%%%%%%%%%%%%%%%%%%%%%%%%%%%%%%%%%%%%%%%%
%\cite{Myers:1986rx}
\bibitem{Myers:1986rx}
  R.~C.~Myers,
  %``Higher Dimensional Black Holes In Compactified Space-times,''
  Phys.\ Rev.\  {\bf D35}, 455 (1987). 
  
%\cite{Breckenridge:1996is}
\bibitem{Breckenridge:1996is}
  J.~C.~Breckenridge, R.~C.~Myers, A.~W.~Peet and C.~Vafa,
  %``D-branes and spinning black holes,''
  Phys.\ Lett.\  B {\bf 391}, 93 (1997). 
  [arXiv:hep-th/9602065].  
  
%\cite{Maeda:2006hd}
\bibitem{Maeda:2006hd}
  K.~-i.~Maeda, N.~Ohta, M.~Tanabe,
  %``A Supersymmetric Rotating Black Hole in a Compactified Spacetime,''
  Phys.\ Rev.\  {\bf D74}, 104002 (2006). 
  [hep-th/0607084].
  
%\cite{Ishihara:2006iv}
\bibitem{Ishihara:2006iv}
  H.~Ishihara, M.~Kimura, K.~Matsuno and S.~Tomizawa,
  %``Kaluza-Klein multi-black holes in five-dimensional Einstein-Maxwell
  %theory,''
  Class.\ Quant.\ Grav.\  {\bf 23}, 6919 (2006). 
  [arXiv:hep-th/0605030].

%\cite{Matsuno:2008fn}
\bibitem{Matsuno:2008fn}
  K.~Matsuno, H.~Ishihara, T.~Nakagawa and S.~Tomizawa,
  %``Rotating Kaluza-Klein Multi-Black Holes with Godel Parameter,''
  Phys.\ Rev.\  D {\bf 78}, 064016 (2008). 
  [arXiv:0806.3316 [hep-th]].  
%%%%%%%%%%%%%%%%%%%%%%%%%%%%%%%%%%%%%%%%%%%%%%%%%%%%%%%%%%%


%%%%%%%%%%%%%%%%%%%%%%%%%%%%%%%%%%%%%%%%%%%%%%%%%%%%%%%%%%%
\bibitem{Bunting:1987} 
  G.~L.~Bunting and A.~K.~M.~Masood-ul-Alam,
  %``Nonexistence of multiple black holes in asymptotically Euclidean static vacuum space-time,''
  Gen.\ Rel.\ Grav.\ \ {\bf 19}, 147  (1987). 
  

\bibitem{Chrusciel:2005pa} 
  P.~T.~Chrusciel, H.~S.~Reall and P.~Tod,
  %``On non-existence of static vacuum black holes with degenerate components of the event horizon,''
  Class.\ Quant.\ Grav.\  {\bf 23}, 549 (2006). 
  [gr-qc/0512041].
%%%%%%%%%%%%%%%%%%%%%%%%%%%%%%%%%%%%%%%%%%%%%%%%%%%%%%%%%%%
\bibitem{Elvang:2007rd}
  H.~Elvang and P.~Figueras,
  %``Black Saturn,''
  JHEP {\bf 0705}, 050 (2007). 
  [hep-th/0701035].

\bibitem{Iguchi:2007is}
  H.~Iguchi and T.~Mishima,
  %``Black di-ring and infinite nonuniqueness,''
  Phys.\ Rev.\ D {\bf 75}, 064018 (2007), 
  [Erratum-ibid.\ D {\bf 78}, 069903 (2008)]. 
  [hep-th/0701043].
 
\bibitem{Izumi:2007qx} 
  K.~Izumi,
  %``Orthogonal black di-ring solution,''
  Prog.\ Theor.\ Phys.\  {\bf 119}, 757 (2008). 
  [arXiv:0712.0902 [hep-th]].
  %%CITATION = ARXIV:0712.0902;%%

%%%%%%%%%%%%%%%%%%%%%%%%%%%%%%%%%%%%%%%%%%%%%%%%%%%%%%%%%%%
%\cite{IPUC-85-1}
\bibitem{IPUC-85-1} 
  G.~Clement,
  %``Solutions Of Five-dimensional General Relativity Without Spatial Symmetry,''
  Gen.\ Rel.\ Grav.\ \ {\bf 18}, 861  (1986). 
%%%%%%%%%%%%%%%%%%%%%%%%%%%%%%%%%%%%%%%%%%%%%%%%%%%%%%%%%%%
  
%\cite{Khuri:1995xq}
\bibitem{Khuri:1995xq}
  R.~R.~Khuri, T.~Ortin,
  %``A Nonsupersymmetric dyonic extreme Reissner-Nordstrom black hole,''
  Phys.\ Lett.\  {\bf B373}, 56-60 (1996). 
  [hep-th/9512178].
  

%\cite{Reall:2002bh}
\bibitem{Reall:2002bh}
  H.~S.~Reall,
  %``Higher dimensional black holes and supersymmetry,''
  Phys.\ Rev.\  {\bf D68}, 024024 (2003). 
  [hep-th/0211290].


%\cite{Kunduri:2007vf}
\bibitem{Kunduri:2007vf}
  H.~K.~Kunduri, J.~Lucietti and H.~S.~Reall,
  %``Near-horizon symmetries of extremal black holes,''
  Class.\ Quant.\ Grav.\  {\bf 24}, 4169 (2007). 
  [arXiv:0705.4214 [hep-th]].
  
 
\end{thebibliography}
\end{document}